\documentclass[twocolumn,showpacs,preprintnumbers,amsmath,amssymb]{revtex4}

\usepackage{graphicx}
\usepackage{dcolumn}
\usepackage{bm}
\usepackage{bbm}
\usepackage[usenames,dvipsnames]{color}
\usepackage{mathrsfs}
\usepackage{xfrac}

\newcommand{\be}{\begin{equation}}
\newcommand{\en}{\end{equation}}
\newcommand{\bea}{\begin{eqnarray}}
\newcommand{\ena}{\end{eqnarray}}

\newcommand{\hbo}{\hbox to 1 true cm {\hfill } }

\begin{document}


\title{ Vortex characterisation of frustration in the 2d Ising spin glass}

\author{Kurt Langfeld$^{a}$}
\author{Markus Quandt$^b$}
\author{Wolfgang Lutz$^{a,b}$}
\author{Hugo Reinhardt$^b$ }

\affiliation{%
\bigskip
$^a$School of Computing \& Mathematics,
Plymouth, PL4 8AA, United Kingdom. \\ }

\affiliation{%
$^b$Institut f\"ur Theoretische Physik, Universit\"at T\"ubingen,
D-72076 T\"ubingen, Germany. \\
}%

\date{ December 16, 2010
}

\begin{abstract}
The frustrated Ising model on a two-dimensional lattice with
open boundary conditions is revisited. A hidden $\mathbbm{Z}_2$ gauge symmetry
relates models with different frustrations which, however, share
the same partition function. By means of a duality transformation, it is shown
that the partition function only depends on the distribution of gauge
invariant vortices on the lattice. We finally show that the exact
ground state energy can be calculated in polynomial time using
Edmonds' algorithm.
\end{abstract}

\maketitle

\paragraph*{Introduction}
Spin glasses~\cite{Binder86} are magnetic materials in which the
magnetic moments are subject to ferromagnetic or anti-ferromagnetic
interactions, depending on the position of the moments within the
sample. The system is \emph{frustrated} in the sense that the
arrangement of spins which minimises the total energy cannot be
determined by considering a local set of spins.  Stated
differently, the change of a single spin might cause a reordering of
many spins when the system relaxes towards a new minimum of
energy~\cite{Alava98}.  Spin glasses undergo a freezing transition to
a state where the order is represented by clusters of spins with mixed
orientations. The relaxation times towards equilibrium are typically
very large, which impedes efficient simulations.

\vskip 1mm
Many efforts have been undertaken to explore equilibrium properties
of spin glasses by means of Monte Carlo
simulations~\cite{Binder79,Binder80,Young82,Ogielski85,Huse85,Rieger96,Matsubara97,Shirakura97}.
Thereby, many insights have been obtained from the simple case
of the 2d Ising model on a square lattice. For the \emph{discrete model},
the bond interactions take values $\pm 1$ at random,
and the model is characterised by the the probability $\kappa $
of finding an anti-ferromagnetic interaction at a given bond.

\vskip 1mm
As first noticed by Bieche et al.~\cite{Bieche80} and further
elaborated by Nishimori~\cite{Nishimori1981,Nishimori1983},
the Ising model with random distribution of anti-ferromagnetic bonds
has a hidden $\mathbbm{Z}_2$ gauge symmetry. As discussed below, this symmetry
implies that gauge invariant observables such as the thermal energy or
the specific heat are unchanged by a certain redistribution of the
anti-ferromagnetic bonds (which may also change considerably in number).
By exploiting this invariance, Nishimori was able to obtain exact results
for special distributions of the bond frustration and
temperature~\cite{Nishimori1981,Nishimori1983}.

\vskip 1mm
In this letter, we further explore the consequences of the hidden
$\mathbbm{Z}_2$ gauge symmetry which relates models with the same partition
function but with different frustrations. We show that within the class of
gauge equivalent models, there is always one model for which the
ground state is homogeneous. Furthermore, we show by means of a
duality transformation that the partition function of any spin glass
only depends on the distribution of gauge invariant vortices.
The amount of frustration in the $2d$ Ising model is quantified by
counting the fraction $\rho$ of \emph{vortices} (non-trivial plaquettes) in a
given bond distribution, and we determine
the exact ground state energy as a function of $\rho$
using Edmonds' algorithm~\cite{Edmonds65a,Edmonds65b}.

\vskip 1mm
\paragraph*{Hidden gauge symmetry}
The partition function of the frustrated Ising model involves a summation over
all spin configurations $\{\sigma_x\}$
\be
Z \; = \; \sum _{ \{\sigma_x\} } \; \exp \Bigl\{ \beta
\sum _{\ell=\langle xy \rangle} U _{\ell} \; \sigma _x \;
\sigma _y \Bigr\} \; ,
\label{eq:1}
\en
where the spins located at the sites of the lattice take the values
$\sigma _x = \pm 1$. We work with open boundary conditions throughout this
paper. The sum in the exponent extends over all bonds
$\ell = \langle xy \rangle$ and the bond variables
$U _{\ell}$ are chosen equal to $+1$, except for a fraction $\kappa$
of randomly distributed bonds where $U_\ell = (-1)$.
In the zero temperature limit, $\beta \rightarrow \infty $, the anti-ferromagnetic
couplings induce frustration.

\vskip 1mm
It was first observed by Nishimori~\cite{Nishimori1981,Nishimori1983}
that bond distributions with vastly different values for  $\kappa $
may still share the same thermodynamical properties. This is due to a
$\mathbbm{Z}_2$ gauge symmetry.
The partition function, eq.~(\ref{eq:1}), and observables such as the thermal
energy,
\be
E(\beta) \; = \; -\frac{\partial \ln Z}{\partial \beta}\; = \; -\;
\Bigl\langle \sum _{\ell = \langle xy \rangle }
\; \sigma _x \, U_\ell \, \sigma _y \; \Bigr\rangle   \; ,
\label{eq:12a}
\en
are invariant under the following change of bonds and spin variables:
\bea
\sigma ^\Omega (x) &=& \Omega (x) \; \sigma (x)
\nonumber \\
U ^\Omega _{\langle xy \rangle  } &=& \Omega (x) \;
U _{\langle xy \rangle  } \; \Omega (y) \; ,
\label{eq:19}
\ena
where the \emph{gauge transformation} takes values in $\mathbbm{Z}_2$,
$\Omega (x) = \pm 1$.

\vskip 1mm
{A simple consequence of this symmetry is that we can always find a gauge}
for which  the ground state is uniform. In particle physics, this gauge is
known as  \emph{Landau gauge}.
It corresponds to the bond distribution with the minimal number of
anti-ferromagnetic bonds:
\be
\sum _{\ell = \langle xy \rangle }
\; U ^\Omega _{\ell } \; \stackrel{\Omega }{
\longrightarrow } \; \hbox{max} \; .
\label{eq:21}
\en
For this choice of bond distribution, the uniform state is a
ground state:
\be
\sigma _x^{\Omega _L} \; = \; \sigma _y^{\Omega _L}  \; = \; \hbox{const.}
\; \; \forall\, x,y \; , \; \;  \hbox{ (Landau gauge) }  .
\label{eq:22}
\en
To see this, we can use eqs.~(\ref{eq:19}) and (\ref{eq:21}) to express
the energy of a given spin configuration  $\{\sigma \}$ in a Landau gauge
background as
\bea
E[\sigma ] &=& - \sum _{\ell = \langle xy \rangle } \sigma _x \; U_{\ell}^{\Omega_L}
\; \sigma _y =  - \sum_{\ell = \langle xy \rangle} U_\ell^{\sigma\cdot \Omega_L}
\nonumber \\
 &\ge& - \sum _{\ell = \langle xy \rangle } U ^{\Omega _L}
_{\ell} =
 - \sum _{\ell= \langle xy \rangle } \sigma _x^{\Omega _L} \;
U ^{\Omega _L} _{\ell} \;  \sigma _y^{\Omega _L} = E[\sigma^{\Omega_L}] \; ,
\nonumber
\ena
where we used the maximum condition eq.~(\ref{eq:21}) for the inequality and
the definition eq.~(\ref{eq:22}) for the uniform ground state, which
satisfies $\sigma _x^{\Omega _L} \sigma _y^{\Omega _L} = 1$, $\forall\, x,y$.
It might turn out that several sets of gauge transformations $\{\Omega \}$
accomplish the (global) maximum in (\ref{eq:21}).
The degeneracy of the spin glass ground state is then twice the number of
these sets. What we have shown is that among these sets there is always
the uniform state.

\smallskip
\paragraph*{Vortex description}
Let us further specify the frustration of the model in a gauge invariant
way. To this end, we introduce plaquette variables,
$U_p \; = \; \prod _{\ell\in p} U_\ell $, constructed from the given bond
background. (The product is along the four bonds forming an elementary
square $p$.) This definition is borrowed from lattice gauge theory, where
a non-trivial value $U_p=-1$ indicates that a $\mathbbm{Z}_2$ vortex intersects
the elementary square $p$. Formally, we set $ V_{p^\ast} = U_p $
for all dual plaquettes $p^\ast$ of dimension $(d-2)$ and locate a vortex
if $V_{p^\ast}=-1$. (In $d=2$, $p^\ast$ is the midpoint of a square $p$.)
With open boundary conditions the plaquettes or vortices determine the bond
variables $U_\ell$ completely, up to $\mathbbm{Z}_2$ gauge transformations.
For this reason, the distribution of vortices is the proper measure to
unambiguously quantify the frustration of the model.

\smallskip
To see this in more detail, we invoke a duality transformation.
We begin by expanding the exponential in (\ref{eq:1}) and use
$[\sigma _x \, U_{l} \, \sigma _y]^2=1$ to find
\be
Z = (\mathrm{cosh} \beta)^{N_\ell}   \; \sum_{\{\sigma _x\}} \;
\prod _{\ell=\langle x y \rangle }
\left[1 +\tanh(\beta)\; \sigma _x \, U_{\ell} \,\sigma _y \right] \; ,
\label{eq:2}
\en
where $N_\ell$ is the number of bonds.
Expanding the products and summing over all configurations of $N$ spins as usual,
the partition function becomes
a sum over all sets of closed loops on the lattice:
\be
Z = 2^N(\cosh\beta)^{N_\ell}   \; \sum_{\mathscr{C}} \;
[\mathrm{tanh} \beta ]^{L(\mathscr{C})} \; W(\mathscr{C}) \; .
\label{eq:3}
\en
Notice that a particular \emph{loop set} $\mathscr{C}$ in the sum may,
in general, contain multiple closed loops that are disconnected, touching or
even intersecting, i.e.~the loops in the set may share points, but not bonds.
Furthermore, $L(\mathscr{C}) = |\mathscr{C}|$
is the total \emph{number} of bonds in the loop set, while $W(\mathscr{C})$,
which is called the \emph{Wilson loop} in gauge field theories,
is their \emph{product}:
\be
 W(\mathscr{C}) \; = \; \prod _{\ell \in \mathscr{C} } U_\ell \; .
\label{eq:4}
\en


The loop formulation eq.~(\ref{eq:3}) is valid in any number $d$ of space
dimensions. It is remarkable that the bond variables $U_\ell$ of the
given spin glass instance only appear in the gauge invariant Wilson loop factor
$W(\mathscr{C})$.
To show that this information is equivalent to the distribution of non-trivial
plaquettes, we only have to appeal to Stokes' theorem
\be
W(\mathscr{C}) = \prod_{p \in A} U_p\,,\qquad\qquad
U_p \equiv \prod_{\ell \in p} U_\ell\,,
\label{eq:5}
\en
where $A$ is any area on the lattice bounded by the loops in
$\mathscr{C}$. Thus, we have
shown that the distribution of non-trivial plaquettes $U_p$ (or vortices
$V_{p^\ast}$) determines the Wilson factor for all loop sets $\mathscr{C}$,
and thus characterises the given spin glass completely via eq.~(\ref{eq:3}).

\smallskip
To complete the gauge invariant duality transformation, we would like to
convert the loop formulation (\ref{eq:3}) to a spin model on the dual
lattice, with the vortices appearing as the spin glass background.
More precisely, we map loop sets $\mathscr{C}$ onto configurations
of spin variables $\tau_{p^\ast} \in \{ \pm 1 \}$ as follows:
\begin{enumerate}
\item start with an empty configuration $\tau_{p^\ast} = 1\quad \forall\, p^\ast$ ;
\item decompose $\mathscr{C}$ in a union of connected loops $\mathscr{C}_i$ ;
\item for each $\mathscr{C}_i$, choose an area $A(\mathscr{C}_i)$
bounded by $\mathscr{C}_i$ ;
\item transform the spin configuration $\{ \tau_{p^\ast} \}$ according to
\be
\tau_{p^\ast} \to \left\{ \begin{array}{l @{\quad ; \quad}l}
- \tau_{p^\ast} & p \in A(\mathscr{C}_i) \\[2mm]
+ \tau_{p^\ast} & p \notin A(\mathscr{C}_i)\,;
\end{array} \right.
\label{trafo}
\en
\item repeat steps 3.~and 4.~until all $\mathscr{C}_i$
have been used.
\end{enumerate}

\noindent
The resulting spin configuration $\{ \tau_{p^\ast} \}$ does not depend on the
choice of loop decomposition in step 2, but it \emph{does} depend on the
subsequent choice of surfaces $A(\mathscr{C}_i)$ in step 3. An exception
is the case $d=2$ where $A(\mathscr{C}_i)$ is the unique area enclosed by the
loop $\mathscr{C}_i$ and the resulting $\{ \tau_{p^\ast} \}$ is unique. On the other hand, the reconstruction of the $\mathscr{C}$ from the dual
spin configuration is unique in any number of dimensions $d$, since
$\{ \tau_{p^\ast} \}$ has the property
\be
\prod_{p^\ast \in \partial\ell^\ast}\tau_{p^\ast} =
\left\{\begin{array}{l @{\quad ; \quad}l}
- 1 & \ell \in \mathscr{C} \\[2mm]
+ 1 & \ell \notin \mathscr{C}\,.
\end{array} \right.
\label{eq:7}
\en
This is because $n_{p^\ast} \equiv \frac{1}{2}\big[1 - \tau_{p^\ast}\big]$ indicates
whether a particular plaquette $p$ is contained an even ($n_{p^\ast} = 0$) or
odd ($n_{p^\ast} = 1$) number of times within the total area $A(\mathscr{C})$.
A bond $\ell$ on the initial lattice is clearly part of the boundary
$\mathscr{C}$ iff the ``area count'' of all plaquettes attached to it is odd,
\[
\sum_{p^\ast \in \partial\ell^\ast} n_{p^\ast} = \sum_{p^\ast \in \partial\ell^\ast}
\frac{1}{2} \big[1 - \tau_{p^\ast}\big] = \mbox{odd}
\]
from which eq.~(\ref{eq:7}) follows using $(-1)^{\frac{1}{2} (1-z)} = z,\, z \in \mathbbm{Z}_2$.
Thus, the map $ \mathscr{C} \to \{ \tau_{p^\ast} \}$ is one-to-many
except for \mbox{$d=2$}, where it is one-to-one.


\smallskip
With the construction of the dual spin variables, the duality transformation
is now simple. First we have (in \emph{any} number of dimensions),
\begin{eqnarray}
L(\mathscr{C}) &=& \sum_{\langle p^\ast q^\ast \rangle} n_{\langle p^\ast q^\ast
\rangle} = \sum_{\langle p^\ast q^\ast\rangle} \frac{1}{2} \,\big[ 1 -
\tau_{p^\ast}\,\tau_{q^\ast}\big] \label{eq:100} \\[2mm]
W(\mathscr{C}) &=& \prod_{p \in A(\mathscr{C})} U_p = \prod_p U_p^{n_{p^\ast}}
= \prod_p U_p^{\frac{1}{2} [ 1 - \tau_{p^\ast} ]} \,. \nonumber
\end{eqnarray}
In $d=2$, the map $\mathscr{C} \to \{ \tau_{p^\ast} \}$ is one-to-one, and
we can directly replace the sum over loop sets by a sum over dual spin
configurations:
\bea
Z_{d=2} &=& 2^{N-\sfrac{N_\ell}{2}}\sinh(2\beta)^{\sfrac{N_\ell}{2}} \times
\label{eq:10} \\[2mm]
&&
\times \sum_{\{ \tau_{p^\ast} \}} \; \exp \left\{ \widetilde{\beta}
\sum_{\langle p^\ast q^\ast \rangle } \tau_{p^\ast} \tau_{q^\ast} \right\}
\; \prod_{p^\ast} \big[\,V_{p^\ast}\,\big]^{ \frac{1}{2}[1-\tau_{p^\ast}]} \; ,
\nonumber
\ena
where $\widetilde{\beta}\equiv - \ln\big(\mathrm{tanh} \, \beta\big)/2$
and $V_{p^\ast} = U_p$ is the vortex background. This formulation
lives entirely on the dual plaquettes, i.e.~the sites of the dual lattice.
For convenience, we can also rewrite the Wilson loop factor as
\be
W \equiv \prod_{p^\ast} \big[\,V_{p^\ast}\,\big]^{ \frac{1}{2}[1-\tau_{p^\ast}]}
= \prod_{p^\ast\,:\, V_{p^\ast} = -1} \tau_{p^\ast}\;.
\label{eq:12}
\en
This shows explicitly that the partition function of the spin glass is entirely determined by the location
of vortices ($V_{p^\ast} = -1$) on the dual lattice.

\smallskip
Finally, we can also rewrite the vortex contribution to the partition function
as an imaginary external magnetic field. \mbox{However}, although the low-temperature limit is attained for
$\widetilde{\beta} \to 0$ suggesting a standard strong-coupling
expansion, the dual theory is still tedious to solve due to the
inherent sign problem.

\smallskip
The extension to $d > 2$ space dimensions is straightforward, with spin
configurations now assigned to dual bonds, plaquettes etc. However, the
transformation (\ref{trafo}) is complicated by the fact that a given loop configuration
$\mathscr{C}$ does \emph{not} lead to a unique dual spin configuration
$\{\,\tau_{p^\ast}\,\}$, because there are many surfaces $A(\mathscr{C})$ over
$\mathscr{C}$ which enter the construction of the dual spins. In a certain sense,
this is a new (geometrical) gauge symmetry which requires a  unique prescription
of how to assign areae $A(\mathscr{C})$ to closed loops in $d>2$. Only such a
fixed partition function is equivalent to the loop formulation eq.~(\ref{eq:3}),
but the ``area fixing'' is presumably NP hard in $d>2$.


%
\begin{figure*}[t]
\includegraphics[width=0.3\linewidth]{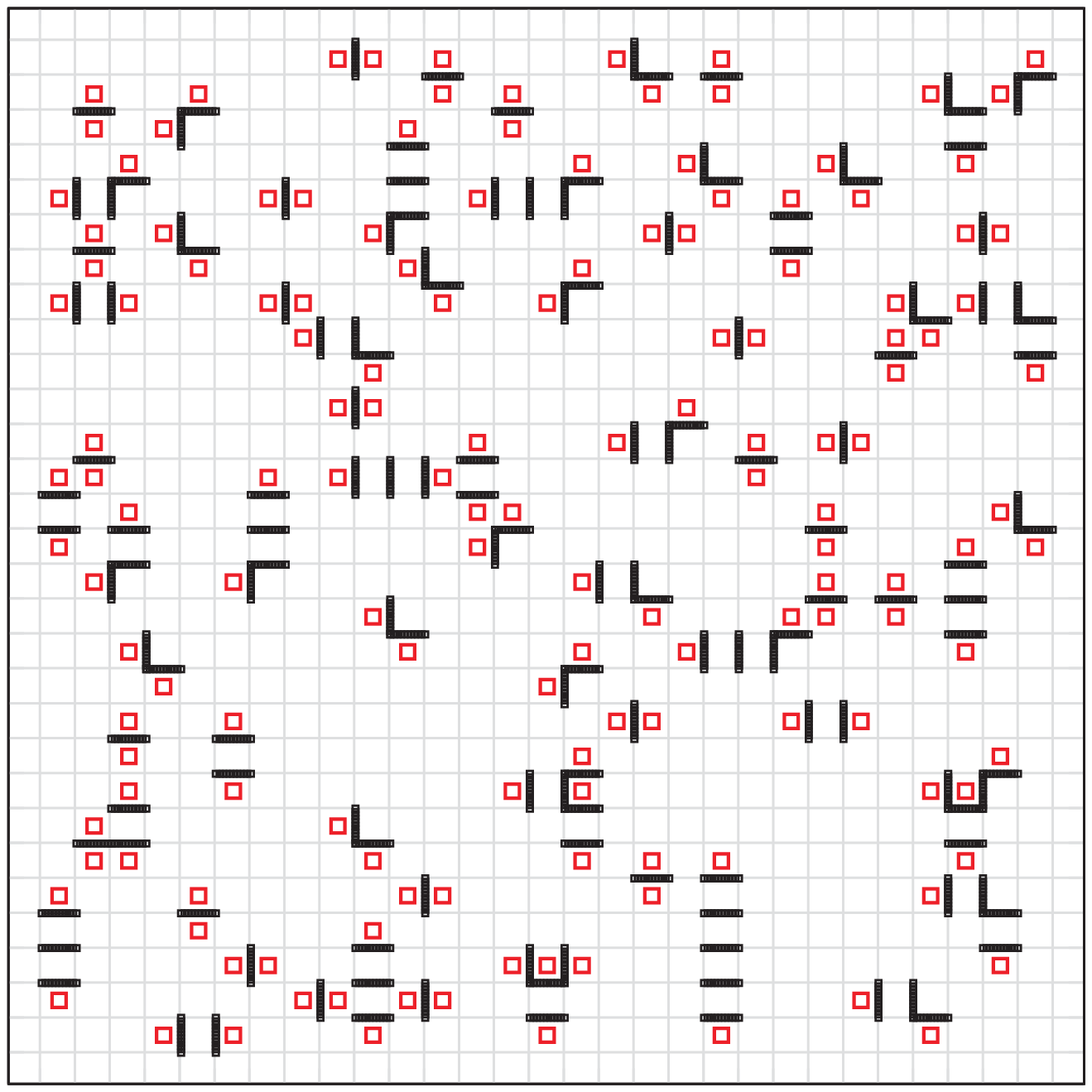} \hspace{3.5cm}
\includegraphics[height=0.3\linewidth]{energy_rho.eps}
\caption{Left panel: $N_V=150$ randomly distributed vortices
(red squares) on a $30 \times 30$ lattice and the corresponding
minimal matching with anti-ferromagnetic bonds (black lines). $E_0/N_\ell \; = \; -0.82759(1)$. Right panel:
Ground state energy per bond as a function of the density of frustration
$\rho$ on $60^2$, $90^2$ and $120^2$ lattices with open boundary conditions.
The two fits refer to eqs.~(\ref{eq:23b}) and (\ref{eq:42}), respectively.
}
\label{fig:2}
\end{figure*}

\smallskip
\paragraph*{Numerical studies}
In $d=2$, the ground state energy can be obtained in polynomial time.
To see this, consider the dual formulation (\ref{eq:10}) in the low-temperature
limit $\tilde{\beta} \to 0$, when the exponential can be expanded. Using the
representation (\ref{eq:12}) for the Wilson factor, the subsequent sum over all
spin configuration gives zero unless every spin variable  $\tau_{p^\ast}$
appears an even number of times, i.e.~all spin variables from eq.~(\ref{eq:12})
have been paired by corresponding variables from the expansion of the exponential.
At order $\tilde{\beta}^1$, the exponential gives one pair of adjacent
$\tau_{p^\ast}$'s which can match two adjacent vortices from the Wilson loop;
at order $\tilde{\beta}^2$, we can match two pairs of adjacent vortices, or
one vortex pair separated by two dual bonds. 
Proceeding in this way, it becomes
clear that the lowest non-vanishing contribution to the partition function
is given by the \emph{minimal matching} of all vortices in the Wilson factor,
\[
Z_{d=2} \simeq 2^{N - \sfrac{N_\ell}{2}}\,\sinh(2\beta)^{\sfrac{N_\ell}{2}}\cdot
\nu\cdot \tilde{\beta}^{\,N_A}\,
\]
where $N_A$ is the length (number of bonds) in the minimal matching, and
$\nu$ is the degeneracy (number of distinct minimal matchings).
The ground state energy thus becomes
\be
E_0 = - N_\ell - N_A \,\lim_{\beta\to\infty} \frac{\partial \ln \tilde{\beta}}{\partial \beta}
= 2 N_A - N_\ell\,.
\label{eq:50}
\en
The core of the solution is therefore the minimal matching of vortex pairs on the
dual lattice, for which \emph{Edmonds' algorithm} \cite{Edmonds65a,Edmonds65b}
gives an answer in polynomial time~\cite{Bieche80}.

\smallskip
Figure~\ref{fig:2} shows a random distribution of $N_V=150$ vortices
on a $L\times L$ lattice with $L=30$ and open boundary conditions, and the
corresponding minimal number of frustrated bonds obtained with Edmonds' algorithm.
Also shown is the ground state energy per bond as a function of the
vortex density $ \rho \equiv N_V / (L-1)^2$.
The data comprise an average over 100 random vortex distributions for each value of
$\rho $. It turns out that the ground state energy per bond is well fitted
for small values of $\rho $ by
\be
E_0(\rho ) / N_\ell \; \approx \; -1 \; + \; 0.39(2) \; \rho ^{0.49(2)} \; .
\label{eq:23b}
\en
Two limits can be obtained analytically: For $\rho =0$, we recover
the ferromagnet with $E_0(0) / N_\ell = -1$. For the other extreme
$\rho = +1$ and for $L$ odd, each plaquette of the lattice carries a vortex.
Each vortex pair is saturated by one frustrated bond. Hence,
$N_A=(L-1)^2/2$ bonds out of $2L(L-1)$ bonds are frustrated so that
\begin{equation}
E_0(1) / N_\ell \; = \;  \frac{ 2 N_A -  N_\ell }{N_\ell }
\; = \; - \; \frac{1}{2} \; - \; \frac{1}{2L} \; .
\label{eq:144}
\end{equation}
Here we obtain a $1/L$ correction to the infinite volume result.
Finally, (\ref{eq:23b}) implies that the derivative $dE_0/d\rho$ is
singular for vanishing vortex density indicating that perturbation theory
with respect to the frustration density is not always justified.

\smallskip
There is a simple way to understand eq.~(\ref{eq:23b}). Suppose that the
$N_V$ vortices are distributed uniformly (without correlations) on the lattice.
Each vortex occupies, on average, an area $\rho^{-1}$, which should be taken
(because of the taxi driver metric) as a rectangle with side length $\rho^{-\sfrac{1}{2}}$.
The \emph{greedy} matching of two neighbouring vortices then has length $\rho^{-\sfrac{1}{2}}$,
and the complete greedy matching of all vortices requires
$N_A \simeq \sfrac{N_V}{2}\cdot \rho^{-\sfrac{1}{2}}$ bonds. Hence we obtain from eq.~(\ref{eq:50}) a ground state energy of
\begin{equation}
\frac{E_0(\rho)}{N_\ell} \simeq -1 + \sqrt{\rho}\,\, \frac{(L-1)^2}{2L(L-1)}
\approx -1 + \frac{1}{2}\cdot \sqrt{\rho}\,.
\label{eq:42}
\end{equation}
This explains the peculiar $\sqrt{\rho}-$behaviour for uniformly distributed
vortices. The true ground state is, of course, based on the \emph{minimal}, not the
greedy matching, i.e.~the prefactor $\sfrac{1}{2}$ in the above estimate must
be multiplied by the ratio $\alpha < 1$ of the bond numbers
in the minimal and greedy matching. At small vortex densities $\rho \ll 1$, the
empirical law (\ref{eq:23b}) suggest $\alpha \approx 0.8$, while the exact result (\ref{eq:144})
implies $\alpha = 1$ at $\rho = 1$.

Both formulae eq.~(\ref{eq:23b}) and eq.~(\ref{eq:42}) give good estimates of the
ground state energy for spin glasses with a \emph{uniform} (uncorrelated) vortex distribution.
To test an extremely \emph{correlated} situation, we place $N_V$ vortices as
$N_V/2$ pairs of exactly one bond distance (dumbells).
Obviously, the greedy matching is also minimal in this case, i.e.~we
have one matching bond per vortex pair, or $N_A = N_V/2$. The exact ground state
energy per link is thus
$E_0 / N_\ell = -1 + N_V / N_\ell = - 1 + \rho/2$ for $L \gg 1$.
By comparision, the estimate~(\ref{eq:42}) has a maximal
deviation of $14.6 \%$ for this configuration, attained at $\rho \approx 0.34$,
while the fit eq.~(\ref{eq:23b}) intended for small $\rho$ has a relative error
of less than $9\%$ for all $\rho < 0.6$ even on the correlated dumbell configuration.
Thus, it seems that the above estimates work reasonably well even for
correlated vortex distributions.

\paragraph*{Conclusions}
In conclusion, we have emphasised the importance of a gauge invariant
classification of frustration. The partition function
was found to depend solely on the distribution of gauge invariant
vortices on the lattice. In two dimensions, the exact ground state
energy was calculated in polynomial time using Edmonds' algorithm.
Using numerical simulations, we find the deviation of the ground state
energy from that in the ferromagnetic case is to a good extent
proportional to the square root of the vortex density.

\smallskip
\noindent {\bf Acknowledgments}
We thank A.K.~Hartmann for useful information.
KL is supported by \emph{STFC}, UK  under contract ST/H008853/1 and
HR is supported by \emph{DFG}, Germany under
contract DFG-Re856/4-2.

\end{document}